\documentclass[12pt]{article}
\usepackage{amsmath}
\usepackage{graphicx}
\begin{document}
\baselineskip=18 pt
\begin{center}
{\large{\bf  An axially symmetric spacetime with causality violation}}
\end{center}

\vspace{.5cm}

\begin{center}
{\bf Bidyut Bikash Hazarika}\footnote{bidyutbikash116@gmail.com}\\
{\it Department of Physics, Gauhati University,}\\
{\it Guwahati-781014, Assam, India}\\

\end{center}

\vspace{.5cm}

\begin{abstract}
We present an axially symmetric spacetime which contains closed timelike curves, and hence violates the causality condition. The metric belongs to type III in the Petrov classification scheme with vanishing expansion, shear and twist. The matter-energy represents a pure radiation field with a negative cosmological constant. The spacetime is asymptotically anti-de Sitter space in the radial direction.

\end{abstract}



\vspace{.5cm}

\section{Introduction}
Time-travel is amongst the wierdiest predictions of the general theory of relativity (GTR). There are numerous solutions of the field equations of GTR that allow for the formation of closed timelike curves (CTCs); trajectories in spacetime that allows a material object to return to a point in its past. The G\"odel solution \cite{Go}, describing a rotating universe, was the first one to be studied that contained CTCs. Van Stockum's solution \cite{Van}, found years before Godel's, was later shown to contain CTCs \cite{Van1}. Some other prominent examples include the Kerr and Kerr-Newmann black holes solution where CTCs form inside the event horizon \cite{Kerr,Kerr1}, the NUT-Taub metric \cite{NUT}, Gott's solution \cite{Gott} of two infinitely long cosmic strings and the Alcubierre warp-drive \cite{Warp} solution. All the above mentioned time machine spacetimes belong to the class of eternal time machines, where CTCs pre-exist. ``True" time machine spacetimes are those where CTCs form at some particular instant of time. The Ori time machine spacetime \cite{Ori}, which is locally isometric to plane wave spacetimes, is a prime example in this category. Some other examples would be the spacetimes discussed in \cite{Faiz,Faiz1,Faiz2,Faiz3,Faiz4,EPJP,Ori1,Faiz6}.

Most of the time machine spacetimes discussed in the literature violate one or more of the energy conditions. The various energy conditions of the general theory with their interpretations are \cite{Step}\\
$(i)$ {\it Weak Energy Condition} (WEC): Local mass-energy density must not be negative.

$(ii)$ {\it Strong energy condition} (SEC): Gravitationally active matter and fields should tend to focus rather than de-focus a collimated beam of light.

$(iii)$ {\it Dominant energy condition} (DEC): Energy must not flow faster than light.

$(iv)$ {\it Null energy condition} (NEC): Stress-energy experienced by a light ray must not be negative.

The WEC is violated in the spacetimes discussed in \cite{Warp,Morris,Morris1,Ever,Ever1}, while the SEC is violated in \cite{Ori2,Ori3,Ori4}. The spacetime discussed in this work satisfies the energy conditions. The matter-energy of the spacetime represents a pure radiation field with a negative cosmological constant.

A spacetime with a negative cosmological constant is termed anti-de Sitter (AdS)-like. The AdS spacetime has huge theoretical implications, particularly after the advent of the AdS/CFT correspondence \cite{Mald} in superstring theory. The AdS/CFT correspondence provides a connection between a quantum theory of gravity on an asymptotically AdS spacetime and a lower-dimensional conformal field theory (CFT) on the boundary of the spacetime.

\section{Analysis of the spacetime} 
  
The axially symmetric metric in $(r, \phi, z, t)$ coordinates is given by
\begin{equation}
ds^2=\frac{dr^2}{\alpha^2\,r^2}+r^2\,dz^2+\left(-2\,r^2\,dt+\frac{\beta\,z\,dr}{r^2}-t\,r^2\,d\phi\right)\,d\phi
\label{temp}
\end{equation}
Here, the constants $\alpha$ and $\beta$ are non-zero real numbers with $\beta>0$. The coordinates are labeled $x^{1}=r$, $x^{2}=\phi$, $x^{3}=z$ and $x^{4}=t$. The ranges of the coordinates are $0 \leq r < \infty$, $-\infty < z < \infty$, $-\infty < t < \infty$ and $\phi$ is a periodic coordinate $\phi\sim\phi+\phi_0$, with $\phi_0>0$. The metric is Lorentzian with signature $(+, +, +, -)$ and the determinant of the corresponding metric tensor $g_{\mu\nu}$,
\begin{equation}
det\;g=-\frac{r^4}{\alpha^2}
\label{det} 
\end{equation}
It is to be noted that the spacetime (\ref{temp}) has a coordinate singularity at $r=0$.

The non-zero Einstein tensor components of the spacetime (\ref{temp}) are
\begin{equation}
G_{\mu}^{\mu}=3\alpha^2,\quad   G_{\phi}^{t}=-\frac{\alpha^2\,\beta^2}{8\,r^6}
\label{eins}
\end{equation}

We have calculated some of the curvature invariants of the above metric (\ref{temp}), and find that they are constant everywhere:
\begin{equation}
R_{\mu\nu\rho\sigma}\,R^{\mu\nu\rho\sigma}=24\,\alpha^4
\label{inv}
\end{equation}
\begin{equation}
R_{\mu\nu\rho\sigma\,;\lambda}\,R^{\mu\nu\rho\sigma\,;\lambda}=0
\label{inv1}
\end{equation}
Thus the presented spacetime belongs to the class of solutions with constant curvature invariants (CCI spacetimes). 

That the spacetime represented by (\ref{temp}) is axially symmetric is clear from the following. Consider the Killing vector ${\bf{\eta}}=\partial_{\phi}$ having the normal form
\begin{equation}
\eta^{\mu}=\left(0,1,0,0 \right )
\label{killi}
\end{equation}
Its covector is
\begin{equation}
\eta_{\mu}=\left(\frac{\beta\,z}{2\,r^2},-r^2\,t,0,-r^2\right)
\label{killi1}
\end{equation}
The vector (\ref{killi}) satisfies the Killing equation $\eta_{\mu\,;\,\nu}+\eta_{\nu\,;\,\mu}=0$. A cyclically symmetric spacetime admits a Killing vector with spacelike, closed orbits. Axial symmetry means that the spacetime contains a non-empty axis of symmetry. This is ensured if the norm of $\eta^{\mu}$ vanishes on the symmetry axis, {\it i.e.} at $r=0$ \cite{Mars1,Mars2,Mac,Carot,Carot1,Wang}. In our case we find that the norm is
\begin{equation}
\eta_{\mu}\,\eta^{\mu}=-r^2\,t
\label{killi2}
\end{equation}
Closed orbits of the above are spacelike for $t<0$ and the norm vanishes on the symmetry axis. However, the metric fails to satisfy the regularity condition or the elementary flatness condition on the axis \cite{Step}. Hence the axis is ill-defined and the spacetime is not locally flat \cite{Wang1}.

The field equations of gravitation are given by
\begin{equation}
G^{\mu\nu}+\Lambda\,g^{\mu\nu}=T^{\mu\nu}, \quad \mu, \nu=1, 2, 3, 4\quad
\label{eins1}
\end{equation}
where the various symbols have their usual meanings.
The energy-momentum tensor of pure radiation field is given by \cite{Step}
\begin{equation}
T^{\mu\nu}=\rho\,\eta^{\mu}\,\eta^{\nu}
\label{eins2}
\end{equation}
where $\eta^{\mu}$ is a null vector defined by
\begin{equation}
\eta^{\mu}=\left(0,0,0,1\right)
\label{time}
\end{equation}
Using eqn.(\ref{eins2}), eqn.(\ref{eins1}) can be stated as
\begin{equation}
R^{\mu\nu}=\Lambda\,g^{\mu\nu}+\rho\,\eta^{\mu}\,\eta^{\nu}
\label{eins3}
\end{equation}
The non-zero component of the energy-momentum tensor
\begin{equation}
T_{\phi}^{t}=-\rho\,r^2
\label{eins4}
\end{equation}
From (\ref{eins}), (\ref{eins1}) and (\ref{eins3}), we get
\begin{equation}
\Lambda=-3\alpha^2
\label{eins5},
\end{equation}
and
\begin{equation}
\rho=\frac{\alpha^2\,\beta^2}{8\,r^8}
\label{eins6}
\end{equation}
Here $\rho$ is always positive and decreases with increase in $r$, and vanishes as $r\rightarrow \infty$, indicating that the presented spacetime (\ref{temp}) is asymptotically AdS radially. The matter-energy of the pure radiation field always satisfy the energy conditions \cite{Faiz6}.

\section{Classification of the spacetime and its physical interpretation}
The presented spacetime belongs to type III in the Petrov classification scheme. To show this, we construct a set of tetrad vectors $({\bf k},{\bf l},{\bf m},{\bf \bar{m}})$ for the metric (\ref{temp}). Explicitly these vectors are:
\begin{equation}
k_{\mu}=\left(0, 1, 0, 0\right)
\label{null}
\end{equation}
\begin{equation}
l_{\mu}=\left(\frac{-\beta\,z}{2\,r^2},\frac{r^2\,t}{2},0,r^2\right)
\label{null1}
\end{equation}
\begin{equation}
m_{\mu}=\frac{1}{\sqrt{2}}\,\left(\frac{1}{\alpha\,r},0,i\,r,0\right)
\label{null2}
\end{equation}
\begin{equation}
\bar{m}_{\mu}=\frac{1}{\sqrt{2}}\,\left(\frac{1}{\alpha\,r},0,-i\,r,0\right)
\label{null3}
\end{equation}
The set of tetrad vectors above is such that the metric tensor for the line element (\ref{temp}) can be expressed as
\begin{equation}
g_{\mu \nu}=-k_{\mu}\,l_{\nu}-l_{\mu}\,k_{\nu}+m_{\mu}\,\bar{m}_{\nu}+\bar{m}_{\mu}\,m_{\nu} \quad 
\label{null4}
\end{equation}
The vectors (\ref{null})---(\ref{null3}) are null vectors and are orthogonal, except for $k_{\mu}l^{\mu}=-1$ and $m_{\mu}{\bar m}^{\mu}=1$. 
Using the above set of null tetrad vectors we calculate the five Weyl scalars, the only non-zero terms being:
\begin{equation}
\Psi_{3}=\frac{i\,\alpha^2\,\beta}{4\,\sqrt{2}\,r^2}
\label{psi}
\end{equation}
and
\begin{equation}
\Psi_{4}=-\frac{\alpha\,\beta\,\left(i+2\,r\,z\,\alpha\right)}{8\,r^2}
\label{psi1}
\end{equation}
Thus the spacetime (\ref{temp}) belongs to type III in the Petrov classification scheme. Type III regions are associated with a kind of longitudinal gravitational radiation \cite{Step}. The null vector $k_{\mu}$ satisfies the geodesic condition $k_{\mu\,;\,\nu}\,k^{\nu}=0$. Moreover, the kinematic parameters of the spacetime, viz., expansion, twist and shear are as given below:
\begin{equation}
{\bf{\Theta}}=\frac{1}{2}\,k^{\mu}_{;\,\mu}=0
\label{exp}
\end{equation}
\begin{equation}
{\bf{\omega}}^2=\frac{1}{2}\,\left(k_{\mu\,;\,\nu}-k_{\nu\,;\,\mu}\right)k^{\mu\,;\,\nu}=0
\label{twist}
\end{equation}
\begin{equation}
{\bf{\sigma\bar{\sigma}}}=\frac{1}{2}\,\left(k_{\mu\,;\,\nu}+k_{\nu\,;\,\mu}\right)k^{\mu\,;\,\nu}-{\bf{\Theta}}^2=0
\label{shear}
\end{equation}
Hence, the spacetime admits an expansion-free, twist-free and shear-free null geodesic congruence.

\section{Spacetime generating CTCs}

An intriguing property of the presented spacetime (\ref{temp}) is that it generates CTCs that appear after a definite instant of time, thus exhibiting time machine-like behaviour, and hence violating the causality condition. Consider closed orbits of constant $r=r_{0}>0$, $z=z_{0}$ and $t=t_{0}$ given by the 1D line element
\begin{equation}
ds^2=g_{\phi\phi}\,d\phi^2=-t\,r^2\,d\phi^2
\label{temp1}
\end{equation}
These orbits are null curves for $t=t_{0}=0$, spacelike throughout $t=t_{0}<0$, but become timelike when $t=t_{0}>0$, which indicates that CTCs are formed at an instant when $t=t_{0}>0$.

One can ensure that the CTCs evolve from an initially spacelike $t=constant$ hypersurface. This can be ascertained by calculating the norm of $\nabla_{\mu}t$, or equivalently by noting the sign of the inverse metric tensor component $g^{tt}$. For the metric (\ref{temp}), we have
\begin{equation}
g^{tt}=\frac{4\,t}{r^2}+\alpha^2\,\beta^2\,\frac{z^2}{4\,r^6}
\label{temp2}
\end{equation}
A hypersurface $t=constant$ is spacelike when $g^{tt}<0$ for $t<0$, null at $g^{tt}=0$ when $t=0$ and timelike when $g^{tt}>0$ for $t>0$. To conform with the above, we choose constant $z$-planes, $z=z_{0}$, where $z_{0}$ equals zero. The spacelike $t=constant<0$ hypersurface can be chosen as the initial hypersurface over which initial data is specified. There is a Cauchy horizon at $t=t_{0}=0$, called the Chronology horizon, which separates the causal past and future in the past directed and a future directed manner. Hence the spacetime evolves from a partial Cauchy surface ({\it i.e.} Cauchy spacelike hypersurface) in a causally well-behaved manner up to a moment, {\it i.e.} a null hypersurface $t=t_{0}=0$, and the formation of CTCs takes place from causally well-behaved initial conditions.

The metric (\ref{temp}) can be regarded as a 4D extension of the 2D Misner space \cite{Mis}. The metric for the Misner space in 2D is given by
\begin{equation}
ds_{Mis}^2=-2\,dT\,dX-T\,dX^2
\label{mis}
\end{equation}
where $-\infty<T<\infty$ and $X$ is a periodic coordinate. The curves $T=T_{0}$, where $T_{0}$ is a constant, are closed since $X$ is periodic. The curves $T<0$ are spacelike, $T>0$ are timelike while the null curves $T=0$ form the chronology horizon, indicating the formation of CTCs at $T=0$.

For constant $r$ and $z$, our metric (\ref{temp}) reduces to the form
\begin{equation}
ds^2=r^2\,\left(-2\,dt\,d\phi-t\,d\phi^2\right)=\Omega\,\left(-2\,dt\,d\phi-t\,d\phi^2\right)
\label{mis1}
\end{equation}
a conformal Misner space, with $\Omega=r^2$ as the conformal factor. Hence, the formation of CTCs in our spacetime is analogous to that of the Misner space in 2D.
\section{Conclusion and outlook}
Our primary motivation in this paper is to highlight causality violation in a solution of GTR. The matter-energy of the axially symmetric spacetime (\ref{temp}) is pure radiation field with a negative cosmological constant. The spacetime, which is asymptotically AdS radially, satisfies all the energy conditions. The metric belongs to type III in the Petrov classification scheme. 

It is shown here that the presented solution has constant curvature invariants. However, the spacetime contains a parallelly propagated (p.p.) curvature singularity \cite{Hawking} at $r=0$. The divergence of the energy density $\rho$ of the pure radiation field (\ref{eins6}) is a signal for the existence of a p.p. curvature singularity. Let $k_{\mu}$ be a tangent vector (\ref{null}) of a null geodesic $\gamma$ and let $\left\lbrace k_{\mu},p_{\mu},n_{\mu},\bar{n}_{\mu}\right\rbrace$ be basis vectors which define a parallelly propagated pseudo-orthonormal frame along $\gamma$ satisfying
\begin{equation}
k_{\mu}\,k^{\mu}=p_{\mu}\,p^{\mu}=0,\quad k_{\mu}\,p^{\mu}=-1,\quad n_{\mu}\,n^{\mu}=\bar{n}_{\mu}\,\bar{n}^{\mu}=1,
\label{one}
\end{equation}
\begin{equation}
k_{\mu}\,n^{\mu}=k_{\mu}\,\bar{n}^{\mu}=p_{\mu}\,n^{\mu}=p_{\mu}\,\bar{n}^{\mu}=n_{\mu}\,\bar{n}^{\mu}=0,
\label{zero}
\end{equation}
\begin{equation}
k_{\mu\,;\,\nu}\,k^{\nu}=p_{\mu\,;\,\nu}\,k^{\nu}=n_{\mu\,;\,\nu}\,k^{\nu}=\bar{n}_{\mu\,;\,\nu}\,k^{\nu}=0
\label{para}
\end{equation}
Eqn.(\ref{para}) is the condition for $\left\lbrace k_{\mu},p_{\mu},n_{\mu},\bar{n}_{\mu}\right\rbrace$ being parallelly propagated along $\gamma$. Note that $l_{\mu}$, $m_{\mu}$, $\bar{m}_{\mu}$ described by (\ref{null1})-(\ref{null3}) are not parallelly propagated along $\gamma$.

For the presented spacetime (\ref{temp}), $R_{\rho\sigma}\,p^{\rho}\,p^{\sigma}$ diverges at $r=0$ because
\begin{equation}
R_{\mu\nu}\,p^{\mu}\,p^{\nu}=\left(\rho\,\eta_{\mu}\,\eta_{\nu}\right)\,p^{\mu}\,p^{\nu}=r^4\,\rho\,k_{\mu}\,p^{\mu}\,k_{\nu}\,p^{\nu}=r^4\,\rho=\frac{\alpha^2\,\beta^2}{8\,r^4},
\label{ricci}
\end{equation}
where we have used the fact that $\eta^{\mu}$ given by (\ref{time}) satisfies $\eta^{\mu}=-r^2\,k^{\mu}$. Furthermore, since the relation $R_{\rho\sigma}\,p^{\rho}\,p^{\sigma}=2\,R_{\mu\rho\nu\sigma}\,p^{\rho}\,p^{\sigma}\,n^{\mu}\,\bar{n}^{\nu}$ holds \cite{Hawking}, one of the pseudo-orthonormal components of the Riemann tensor blows up at $r=0$ and hence it is a p.p. curvature singularity.

\end{document}